# Realtime observation of a tungsten-promoted size regulation mechanism in a rhodium catalyst at atomic resolution

Petra Specht[1], Joo H. Kang[2], Kartick Tarafder[4], Robert Cieslinski[2], David Barton[2], Bastian Barton[5], Anna Carlsson[3], Lin-Wang Wang [4], Christian Kisielowski[6]

1) Department of Material Science & Engineering, UC Berkeley CA 94720 / USA

2) The DOW Chemical Company, Midland MI 48667 / USA

3)  FEI Company, P.O. Box 80066, KA 5600 Eindhoven, The Netherlands

4) Material Sciences Division, Lawrence Berkeley National Laboratory, One Cyclotron Rd., Berkeley CA 94720 / USA

5) Institute of Physical Chemistry, Gutenberg University Mainz, Welderweg 11, D-55099 Mainz, Germany

6) The Molecular Foundry and Joint Center for Artificial Photosynthesis, Lawrence Berkeley National Laboratory, One Cyclotron Rd., Berkeley CA 94720 / USA

**Abstract**

The static and genuine structure of small rhodium and rhodium/tungsten nanoparticles on an alumina support can be imaged with atomic resolution even if single digit atom clusters are investigated. Low dose rate electron microscopy is key to the achievement and can generally be applied to investigate any similar material. In such conditions it becomes feasible to identify the chemical composition of nanocrystals from quantitative contrast analyses alone by counting atoms. The ability to fully characterize an unaltered, initial state of the objects allows targeting structural excitations or conformational changes induced by the electron beam itself. For the specific case of catalytic Rh:W particles we stimulate a tungsten-promoted size regulation mechanism in real time that is driven by Oswald ripening and can be understood by a strong binding of tungsten atoms to the oxygen atoms of the support, which builds up strain as the cluster sizes increase.

**Introduction**

Extraordinary progress was made designing the static structure of nanoparticles at the atomic scale that enables and improves a rapidly growing number of physical or chemical processes [1-8]. Nonetheless, it is also recognized that there is plenty of room to better understand the transition from static to dynamic behavior, which leads to functionality. Catalytic activity, deactivation and selectivity at the atomic level are examples for processes where both the static and the dynamic of small objects must be understood equally well in order to come to reliable scientific conclusions [9, 10].

While aberration-corrected transmission electron microscopy is a unique method for studying bulk materials with atomic resolution, the analysis of small nanoparticles (≲ 5 nm) with element specificity has been hampered because beam-sample interactions actively alter their pristine structure in an uncontrolled manner. In representative conditions the deposited energy by the high-brightness electron beams that are needed to visualize single atoms or atom columns simply exceeds the total binding energy of small particles [11]. Using a low dose rate method [12] we show here that electron microscopy enables now studying the transition from static to dynamic behavior in real time because the underlying dose fractionalization allows for the recording of pristine structures and specific choices of dose rates allow targeting object excitations. Thereby, initial and transient states of the observed system can be recorded. Combined with ab-intio calculations the approach enables a path to understand the time evolution of nanostructures at the atomic scale.

This study focuses on bimetallic Rh:W catalysts that emerged some time ago and became of renewed interest for the production of ethanol from biomass-derived synthesis gas [13-15]. It is well established that the addition of the tungsten to the rhodium catalysts enhances conversion rates and selectivity for the catalytic production of oxygenated hydrocarbons [13]. Nevertheless, the underlying mechanisms that lead to performance improvements remain debatable since it is unsettled which sites the W atoms occupy in the Rh lattice or what their function may be. We utilize tungsten promoted rhodium catalysts to demonstrate that their genuine structures can now be analyzed even if single-digit atom clusters are considered and that controlled electron beam-induced system excitations stimulate Ostwald ripening that in turn triggers a promotor-driven size regulation mechanism.

**Experimental details**

Rhodium catalysts were prepared by impregnation of alumina with an aqueous solution of rhodium chloride and ammonium metatungstate to yield 2 wt % rhodium and rhodium to tungsten ratio to be 1. The impregnated catalyst was reduced in hydrogen at 300$^\circ$ C and dispersed on holey carbon grids for observation. Related selectivity and conversion measurements are depicted in Figure 1.

The TEAM 0.5 electron microscope is used for this study [16] because it exploits recent technological progress that allows for dose rate variations at different acceleration voltages to produce in-line holograms with single atom sensitivity at atomic resolution [17]. In brief, a Nelsonian illumination scheme [18] controls the electron emission in the high-brightness gun region and, thereby, the sample illumination cone in a broad-beam mode. Dose rates can be chosen as low as a few atto Amperes per Å$^2$ and electrons are only delivered to sample areas that are also imaged by the CCD camera. Unlike other approaches [19], these modifications of the condenser/gun system eliminate uncontrolled object alterations outside the field of view and allow maintaining a static object structure during observation even if the object consist of only a few atoms as shown here. It also allows for a well-defined increase of dose rates that is known to stimulate physically relevant object conformations [20, 21].

Instructive examples are given in Figure 2: Single low dose rate images of a Rh catalyst exhibit hardly any contrast (Fig. 2a) but amplitude and phase images of an electron exit wave function can be restored with outstanding contrast from focus series that contain ≤ 100 low dose rate images recorded in 170 seconds of exposure time (Figure 2b, 2c). The Figure 2d) shows a cluster of 7 Rh or W atoms on alumina. A commonly used high dose rates of 5000 eÅ$^2$/s results in the displacement of single atoms from their equilibrium position. The resulting image blur is electron beam-induced and prohibits a proper atom count. In contrast, reduction of the dose rate allows acquiring a static image where a central atom is seen with six next neighbors (Fig. 2e). Thus, atom clusters that only contain < 10 atoms can be imaged in their pristine, unaltered configuration. Note the presence of an existing dose rate dependence because the accumulated dose of the reconstructed wave function from 20 images in Figure 2e) matches the electron dose in any single image of Figure 2d) (one second exposure time) but only the atom configuration in Fig. 2d) is static.

The electron exit wave functions, which are in-line holograms, are reconstructed from focal series of images using a Gerchberg-Saxton algorithm [22] that is implemented in the McTempas software package [23]. This software package is also used for image processing and an extraction of column positions with pm precision. It is further employed to calculated the phase images by computing focus series of images using multislice simulations that were successively reconstructed to recover exit wave functions with different occupations of the atom columns with Rh and W atoms to match the experiments.

In order to extract chemical information from each column, we combine phase and amplitude of the in-line hologram in the following way: first, the complex electron exit wave function (EW) is propagated by roughly ± 80 Å around gaussian focus in 1 Å steps by multiplying the EW with the wave propagator $\exp(i\pi f k^2)$ in Fourier space (f: defocus, k = |k|: spatial frequency). In a complex plane this propagation traces part of a circle (Figure 3a). Next, the average modulus of a $(0.7 \text{ Å})^2$ area at all column positions is plotted at each propagation step for each column. Polynomial fits of the resulting sine plots yield their maxima and amplitudes with sub-Å accuracy (Figure 3b). While the position of the maximum defines a zero focus plane $f_0$ for each column, the sine amplitude corresponds to the diameter of the defocus circle that can be represented in an Argand plot as shown in Figure 3c). Since the defocus is a wave aberration, it only affects the phase of the scattered part of the EW. Consequently, the defocus circle diameter (DCD) is independent of the reference wave, and equal to the modulus of the interaction wave EW - 1. The latter is in turn proportional to the projected scattering potential, or the weighted sum of atomic numbers as commonly described by an Argand plot [24]. Effectively, the described algorithm is a transform (arg(EW), |EW|) → (DCD, $f_0$) performed at each column position, with the advantages of easy experimental access while the DCD and the zero focus plane $f_0$ are directly related to chemical composition and exit surface roughness, respectively. They provide column mass and object shape at atomic resolution.

Does rates affect significantly our ability to extract chemical composition from image contrast. It was recently demonstrated that the electron beam induces collective atom displacements (phonons), which yield vibrations with amplitudes that are significantly larger than one would expect from Debye Waller factors because their distribution is not Gaussian but skewed and increases logarithmically with dose rates [12]. The Figure 4a) shows this effect for the Rh:W catalysts, where a reduction of the image contrast (or column mass) by a factor of ~2 is measured for a dose rate increase by ~ 2 orders of magnitude. If the signal loss is described by a Gaussian damping function that captures electron beam-induced vibrations one adds mean atom displacements in the range of 45pm to 65 pm that cause significant blur [12]. In Figure 4b) such values are compared with the information limit resolution of the TEAM 0.5 microscope that is set by focus and image spread [24]. Since beam induced atom displacements contribute to image blur they contribute effectively image spread and the native instrument resolution cannot longer be reached. Therefore, a superior image contrast is achieved in recordings with low dose rates that originates from a better effective resolution. In addition, his contrast improvement necessarily improves on a chemical identification of elements from contrast measurements alone.

Density functional theory (DFT) under the generalized gradient approximation (GGA) [24] is applied to calculate the rhodium / tungsten systems. Projector-augmented wave pseudopotentials [26] are used as implemented in the Vienna Ab initio Simulation Package (VASP) code [27]. Calculated energy values and relaxed atomic structures are shown in Figure 8c and in Table I.

**Results**

The Figure 5 compares atomically resolved images of the Rh and Rh:W catalysts deposited on alpha-alumina. In case of the pure Rh catalyst (Figure 5a), the particles are mostly multiple-twinned, three-dimensional objects and occasionally rafts. The alumina substrate flakes are randomly oriented and commonly oriented along high-indexed zone axes towards the electron beam. Occasionally, unidirectional orientation relationships between the substrate and single phase particles can be directly observed as seen in Fig. 5a) for the Rh [100] cube where the 0.134 nm narrow (022) plane spacing of the rhodium matrix aligns with the 0.16 nm narrow (116) plane spacing of the alpha-alumina substrate (inset). Necessarily, a large lattice mismatch causes strain if the particle attachment to the substrate is fully or partly coherent. Strain relaxation can occur along different paths. In two dimensional thin films, a most common mechanism includes the formation of misfit dislocations if critical thicknesses are exceeded [28] but little is known about strain relaxation effects of nm-sized catalysts attached to substrates. As to the investigated Rh particles, their average size can be approximated by a log-normal distribution with an average of (3.08 ± 0.09) nm and a standard width of 0.4 nm (Figure 6a).

The addition of tungsten to the rhodium catalyst leads to a significant change in structure and distribution as shown in Figure 5b). Now the catalysts are composed of rafts and small atom clusters. In many of the observed cases, the detected Rh:W particles are nothing but clusters with single-digit atom numbers that are directly attached to the alumina lattice as shown in the inset of Figure 5b). The related log-normal particle distribution function is shown in Figure 6a). Our Rh:W catalysts exhibit a reduced average particle diameter of (1.75 ± 0.09) nm and a standard width of 0.6 nm.

Their pristine structure is critically sensitive to the rate of incident electrons during imaging [12], but can be preserved with dose rates of ~ 200 e/Å$^2$s or less using an acceleration voltage of 80 kV (Figure 2). In fact, such low dose rates approach values that are otherwise used in biological sciences to acquire electron micrographs of radiation-sensitive organic macromolecules at resolutions of several Å. In contrast, in-line holograms reconstructed from focal image series can deliver sub-Å resolution with minimal specimen damage, exploiting the fact that electron beam-induced object excitations are commonly reversible [12].

The mass of each atomic column is measured as described in Figure 3 and provides the occupation of each column with a well-defined number of atoms. The Figure 6b) shows a histogram of such quantitative column mass measurements from four different experiments that were executed with low dose rates between 26 e$^-$/Å$^2$s through 94 e$^-$/Å$^2$s. The histogram exhibits discrete peaks, each representing a distinct number of rhodium and tungsten atoms in a column. Both elements can be differentiated because the signal difference between a single rhodium atom (0.140 ± 0.023 a.u) and a tungsten atom (0.210 ± 0.023 a.u.) equals 0.070 a.u. and exceeds the 2*sigma error of 0.046 a.u. of the measurement. This cannot be achieved for high dose rates, where sample stability is compromised and column mass signals are reduced by a factor of ~2, as shown in Figure 4. Moreover, the Figure 6b), reveals that the ratio of single Rh and W atoms is ~ 1:2 (15:27), which suggests that single atoms on the alumina substrate are dominantly tungsten atoms. The catalysts themselves are typically ~ 2-4 atoms thick and, therefore, rafts. Two of the analyzed particles are shown in Figures 7 a) and 7 c). Using image analysis together with the simulation procedures described above, we capture lateral column positions with pm precision [29] in the image plane and match the column contrast by their occupation with a suitable number of rhodium and tungsten atoms. These simulations yield the calculated images shown in Figure 7 b) and 7 e) together with the underlying models in Figures 7 c) and 7 f). This unique procedure maps the chemical composition of the two $Rh_xW_y$ catalysts column-by-column, while maintaining structural integrity at a resolution around one Ångstrom during the imaging process with high-energetic electrons. It is now directly visible that the tungsten atoms are randomly distributed in the projected images of the catalysts and their chemical compositions can be determined by simply counting the number of atoms to be $Rh_{0.75}W_{0.25}$ and $Rh_{0.65}W_{0.35}$, respectively. Thermodynamically, one expects the formation of a eutectic $Rh_{0.75}W_{0.25}$ phase [30]. Thus, the particles are thermodynamically stable alloys with a tendency to preferentially bond tungsten atoms to the substrate.

It is of interest to understand the mechanism defining the size of the tungsten promoted rhodium particles, and thereby altering their activity and selectivity. For this purpose, we captured focus series recorded with different dose rates to stimulate system changes. It is noted that the acquisition of such focus series also captures the time evolution of the system as it is constantly irradiated. The Figure 8a) compares the low dose rate (190 e$^-$/Å$^2$s) phase image of Figure 7d) with the phase image of Figure 8b) where a larger dose rate of 2,720 e$^-$/Å$^2$s was used during a successive recording of 61 images from the same particle. The second, high dose rate reconstruction reveals an increase in particle size as compared to the first, caused by local heating and subsequent Ostwald ripening during acquisition [31,32]. Unexpectedly, a further acquisition of images ( > #61) results in an abrupt disintegration of the catalyst as shown in the image series of Figure

8 b'). The created debris sticks to the same location but is less stable in the electron beam because of its reduced particle size [11]. The full image sequence of this disintegration event is showcased in the supplementary movie "RhW-rupture.mov".

Independent of the composition analysis by contrast interpretation, chemical maps are produced by Energy Dispersive Spectroscopy (EDS) close to atomic resolution as described in the supplementary information. They confirm that tungsten is indeed preferentially attached to oxygen atoms of the alumina substrate and partly integrated into the Rh:W particles [S1]. However, electron beam-induced Ostwald ripening [11,31] occurs during the acquisition of the EDS spectra (Figure S2), in stark contrast to our low dose rate approach, which is free of such artifacts. An average $Rh_{0.41}W_{0.59}$ composition was measured close to the intended 1:1 loading of the alumina substrate with these elements. Consistent with our quantitative contrast measurements the EDS also shows that isolated tungsten and rhodium atoms exist on the substrate (inset in Figure 5 b).

**Discussion**

Geometrical distortions, lattice parameters and strain are measured by analyzing diffraction spots (g-vectors) in the Fourier transform of exit wave functions, which are nano diffraction patterns. The Figure 9 shows real and reciprocal space images for the [100] Rh:W raft together with a table that compares the measured diffraction spots with expectations for the monometallic Rh and W metals [33]. The hexagonal alumina substrate is imaged in a high indexed [5-21] direction (= (4,3,-5) plane). Thus, this specific Rh:W particle of Figure 7d) is imaged along a [100] direction with smallest recorded diffraction vectors g of 1/1.9 Å$^{-1}$ and 1/2.1 Å$^{-1}$. These spacings are compatible with either the tungsten (011) planes of 1/2.0 Å$^{-1}$ or with the rhodium (002) planes of 1/1.9 Å$^{-1}$ (Table in Figure 9). However, a Scherrer peak broadening of the diffraction spots ori-ginates from the small particle size and a rhombohedral catalyst distortion exists that prohibit precise composition measurements using Vegard's law. Certainly, a chemical composition of $Rh_{0.65}W_{0.35}$ as suggested by our contrast analysis is compatible with the measured plane spacing. Furthermore, the orientation of (2,0,-10) alumina planes aligns within ~8 degrees with the (022) lattice planes of the tungsten doped rhodium alloy, but the large mismatch in lattice spacings must cause significant strain. We will use this geometry for our ab initio calculations.

We use First Principles density functional theory calculations using general gradient approximations to test the hypothesis that strain energy can be responsible for the rupture of tungsten promoted rhodium particles. For this purpose we studied total and adhesion energies of atom clusters containing 25 and 50 atoms of either Rh or $Rh_{0.76}W_{0.24}$ that are bonded to a non-polar terrace of the (4,3,-5) alumina surface, which is the situation detailed in Figure 9. Following exactly the TEM results and cluster orientations as shown in Figure 9, we use clusters consisting of two (001) metal layers of the FCC crystal structure with their (110) planes aligned with the (2,0,-10) planes of the alumina. After relaxation of the rhodium-only $Rh_{25}$ clusters the average distance between a Rh atom at the bottom of the cluster and the surface of the alumina is rather large, 2.61 Å (Fig. 8c). There is almost no direct covalent bonding between bottom Rh atom and surface O atoms and there is little tendency to align them together and the whole $Rh_{25}$ cluster appears rigid, afloat on top of the alumina surface. For the W promoted catalyst cluster, we locate the W atoms in the bottom layer of the cluster. In contrast with the pure Rh case, a $Rh_{19}W_6$ cluster separates from the substrate by only ~1.98 Å (Fig. 8c). This is due to a strong bonding between W and O atom, and furthermore a W-W pair at the bottom of the cluster can locally match a similar O-O pairs on the substrate for their distances to form strong W-O bonds. However, as the cluster size increases, this local alignment between W-W and O-O pairs will be vi-olated due to the lattice mismatch. As a result, a large cluster catalyst prefers to break down into smaller ones in order to re-establish the local alignment of the atom pairs. The calculated energies shown in Table 1 support this view. When a $Rh_{38}W_{12}$ breaks into two $Rh_{19}W_6$ clusters the total energy of the cluster will increase due to the bond breaking in the cluster. However, its binding energy to the surface increases even more (due to the possibility W-O pair alignment), resulting in a slightly smaller total energy. In contrast, for a $Rh_{50}$ cluster breaking into two $Rh_{25}$ clusters, this binding energy increase is absent. As a result the $Rh_{50}$ cluster has a 1.2 eV lower energy than the two $Rh_{25}$ clusters and there is no tendency to break up.

In summary, it is shown by a quantitative contrast analysis in combination with computation that tungsten atoms alloy with rhodium atoms close to thermodynamic expectations ($Rh_{0.75}W_{0.25}$), but exhibit a preference to strongly bond to oxygen atom pairs of the alumina substrate. Ostwald ripening is induced by the electron beam, which is observed in time-resolved experiments with single atom sensitivity. We find a limitation to the ripening process, since particles disintegrate when a critical diameter of ~3 nm is exceeded. First Principles calculations suggest that this rupture of tungsten rich $Rh_{0.75}W_{0.25}$ particles is driven by the strong bonding of the tungsten atoms to the alumina, which increases the strain energy of the system. The resulting size and element distribution affects catalytic selectivity and conversion.

Outgoing we conclude that our approach allows us studying at atomic resolution the static, pristine structure of small particles and their transitions to dynamic behavior induced by electron beams or thermally stimulated.

**Acknowledgement**

Electron Microscopy was performed at the Molecular Foundry, NCEM, using the TEAM 0.5 microscope which is supported by the Office of Science, Office of Basic Energy Sciences, of the U.S. Department of Energy under Contract No. DE-AC02-05CH11231. The Dow Chemical Company supported CK and PS for the investigations of the Rh, RhW catalysts. Wang is supported by the Material Theory project funded by Office of Science, Office of Basic Energy Sciences, of the U.S. Department of Energy under Contract No. DE-AC02-05CH11231. The computations used computational resources of the National Energy Research Scientific Computing Center and Oak Ridge Leadership Computing Facility with computational time allocated by the Innovative and Novel Computational Impact on Theory and Experiment project.

**Figure Captions**

Figure 1:

Measurements of sensitivity and conversion for the production of oxygenated hydrocarbons from synthesis gas with tungsten and rhodium/tungsten bimetalic catalysts.

Figure 2:

Principles of low-dose rate microscopy.

a) A single low dose rate image where contrast is overwhelmed by noise.

b), c) Amplitude and phase of an in-line hologram. The exit wave function is reconstructed from a focus series of 21 images and comprises an in-line hologram.

d) Single images documenting the time evolution of an atom cluster formed by 7 tungsten atoms. Their motion causes blur and is stimulated by the large dose rate listed below the images, which is representative for a traditional acquisition of high resolution images.

e) A static phase image of such atom clusters can be captured by in-line holography in low dose rate conditions. Note that the accumulated electron dose equals the accumulated dose used to acquire any of the images in figure S2d) during the chosen second of exposure time.

Figure 3:

Quantitative procedures extracting chemical information from electron exit wave functions.

a) Argand plot ( $\psi = A(x,y) e^{i\varphi(x,y)}$, $A(x,y)$ = local amplitude, $\varphi(x,y)$ = local phase) of two arbitrary selected atomic columns averaged over an area of $(0.7\ \text{Å})^2$ around their phase maxima.

Smooth circles segments are created by wave propagation between -80 and +80 Å of defocus in 1024 steps.

b) Moduli of the wave amplitudes plotted as a function of defocus (= propagation distance). The amplitude of the sine function equals the defocus circle diameter (DCD) and is independent of exit and entrance surface roughness. It represents a robust measure of the projected column mass.

c) Schematic relation of DCD measurements to the description of electron exit wave functions in a channelling model [24]. The column mass (number & sort of atoms / column) can be expressed by a phase changes per atomic column or equivalently by DCD values. They relate by the constant 1.5 rad/a.u.(DCD). The experimentally determined phase changes for one Rh atom (0.2 rad / 1.4 a.u) and one W atom (0.3 rad / 2.1 a.u.) is highlighted in the Argand plot that also shows the effect of a small absorption < 10 %.

Figure 4:

Dependence of the contrast measurements on dose rates.

a) Reconstructed phase images of the same nanoparticle recorded with the listed dose rates. An analysis of the data in terms of column mass reveals the depicted contrast reduction with increasing dose rates. The contrast degradation is caused by electron beam-induced object excitations that soften the average scattering potential.

b) Calculated information limit of electron microscopes for different focus and image spreads [24] at 80 kV. The information limit of the TEAM 0.5 microscope is pointed out [12]. The data are compared with electron beam-induces atom vibrations of 45pm amd 65 pm that were measured using the listed dose rates. Electron beam-induced sample excitation prohibit reaching the optimum resolution in this case.

Figure 5:

Characteristic images of the investigated materials. The scale bar is 3 nm in all cases.

a) Single lattice image from Rh catalysts on alpha-alumina. The power spectrum (inset) shows an existing orientation relationship between the highlighted (box) rhodium cube and the substrate.

b) Phase of the electron exit wave function of Rh:W catalysts. The image is reconstructed from a series of low dose rate images. The inset shows a single, high dose rate image of tungsten atoms that are attached to the lattice planes of the alumina substrate. In high dose rate conditions their location changes from image to image.

Figure 6:

Quantitative contrast evaluations.

a) Approximation of the particle size distributions by log-normal distribution functions for Rh catalysts and W promoted Rh catalysts. Average sizes: Rh - (3.08 ± 0.09) nm, Rh:W - (1.75 ± 0.09) nm

b) Histogram of column mass values derived from propagation of the particle exit wave function. The indicated number of atoms is determined experimentally and confirmed by image simulations. A single Rh atom contributes with (0.14 ± 0.023) arbitrary units (phase change of 0.20 ± 0.035 rad/Rh atom in the exit wave function) to the column mass and a single tungsten atom with 0.21+0.023 [a.u.] (p.c. 0.30 + 0.035 rad / W atom). The column mass is evaluated as the diameter of the defocus circle (DCD) occurring in the complex plane upon wave propagation, having the same dimension as the modulus of the wave (Figure 3). Accumulation points occur as multiples of the single atom mass contrasts and are mixed in Rh and W columns. A total of 433 columns and 882 atoms are analyzed with error characteristics as indicated.

Figure 7:

Quantitative analysis of atom column positions and their contrast.

a-c) A Rh:W [110] catalyst at the edge of the substrate is shown over vacuum.

d-f) A Rh:W [100] catalyst on a [5-21] alumina substrate is shown.

a) and c) are phase images of in-line holograms reconstructed from low dose rate images < 100 eÅ2s.

(b) and (e) are simulated phase images of both particles using the McTempas multislice code to simulated image series and compute the in-line holograms by exit wave reconstruction.

(c) and (f) are the underlying models that match column position and contrasts and determine the chemical composition of the catalysts as indicated.

Figure 8:

Time-resolved observation of Ostwald ripening and particle disintegration together with models explaining the occurrence of particle rupture.

a) Phase image of the genuine particle reconstructed from 35 low dose rate images.

b) Phase image reconstructed from 61 images of the particle during beam-stimulated particle growth. The particle size reaches a 3 nm diameter significantly exceeding the average particle size of 1.8 nm.

b') Single images #60 - #63 of the image series that captures the disintegration event. The image series is shown in the supporting movie "RhW-rupture.mov".

c) Models of Rh25, Rh19W6 clusters on terraces of the (4 3 -5 ) alumina plane containing 25 atoms each. Strong W-O interactions reduce the cluster-substrate distance significantly as indicated. This strong interaction breaks up particles as they grow because lattice mismatches cannot be maintained.

Figure 9:

Geometrical relations of the Rh:W [100] raft and the alumina support.

a) Phase image of the low dose rate in-line hologram. Orientation relationship between the substrate and catalyst are indicated. The color code highlights in yellow columns that contain tungsten atoms.

b) Nano diffraction pattern obtained by a Fourier transform of the hologram with characteristic diffraction (g) vector spacing as indicated.

The table compares the measured lattice spacing (Fourier Transform) with the expected spacing of pure rhodium and tungsten metals for selected planes.

Table 1:

Calculated energy values for Rh and RhW atom clusters on terraces of the (4 3 -5 ) alumina plane. One $Rh_{50}$ cluster binding to the substrate is more stable than two $Rh_{25}$ clusters binding to the substrate. However, two $Rh_{19}W_6$ cluster binding to the substrate tend to be more stable than one $Rh_{38}W_{12}$ cluster binding to the substrate. Relaxed model structures are shown.

| Energy / eV | Rh$_{25}$ | Rh$_{50}$ | Rh$_{19}$W$_6$ | Rh$_{38}$W$_{12}$ |
|---|---|---|---|---|
| total cluster | -137.539 | -278.163 | -174.007 | -356.119 |
| adsorption energy on alumina | -8.692 | -15.523 | -12.744 | -17.367 |
| Cluster + adsorption for 50 atoms | 2*(-146.231) =-292.462 | -293.686 stable | 2*(-186.751) =-373.502 stable | -373.486 |
| Relaxed models plane view | 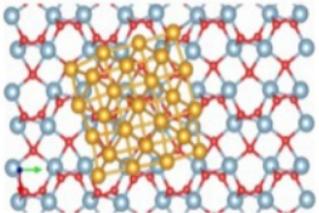 | 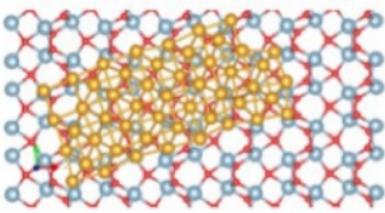 | 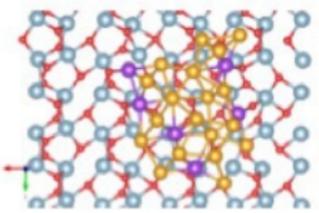 | 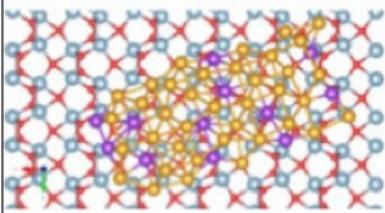 |

Rh&W

Rh only

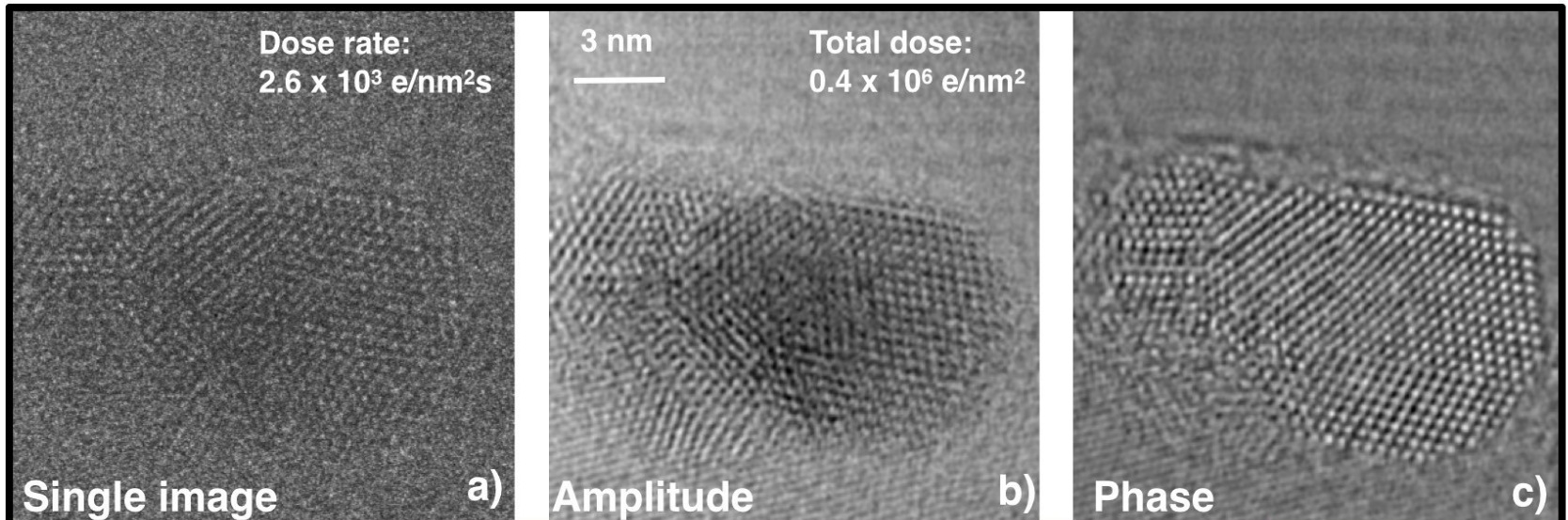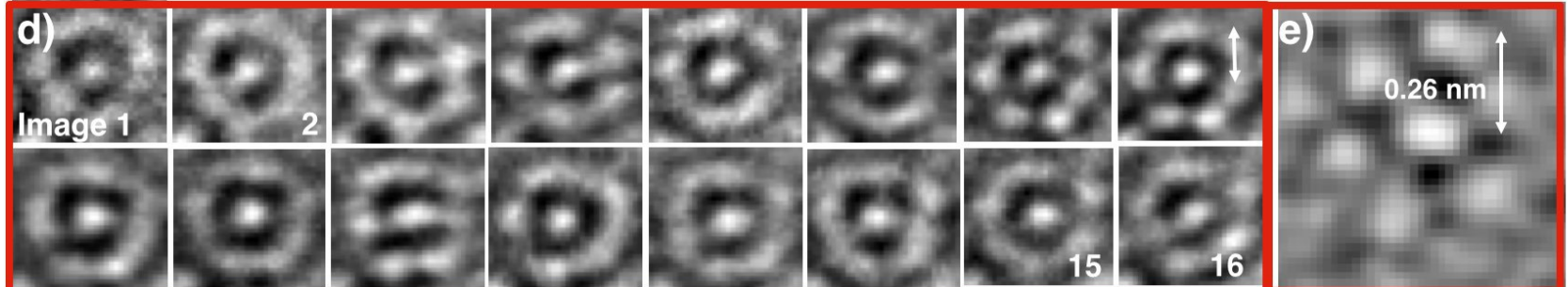

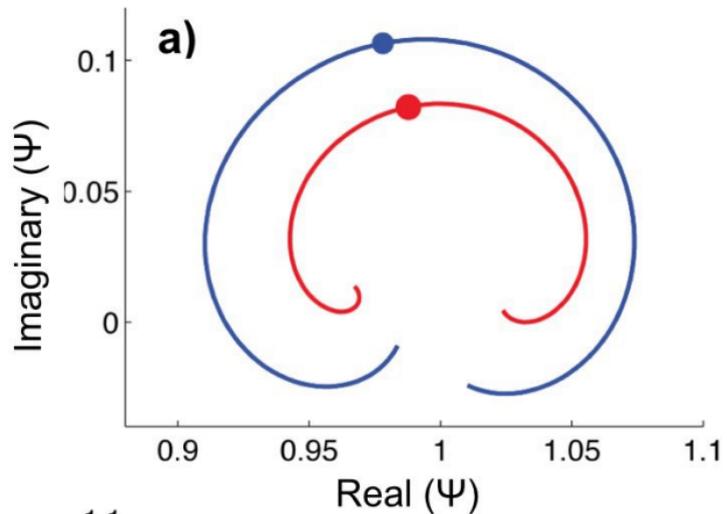
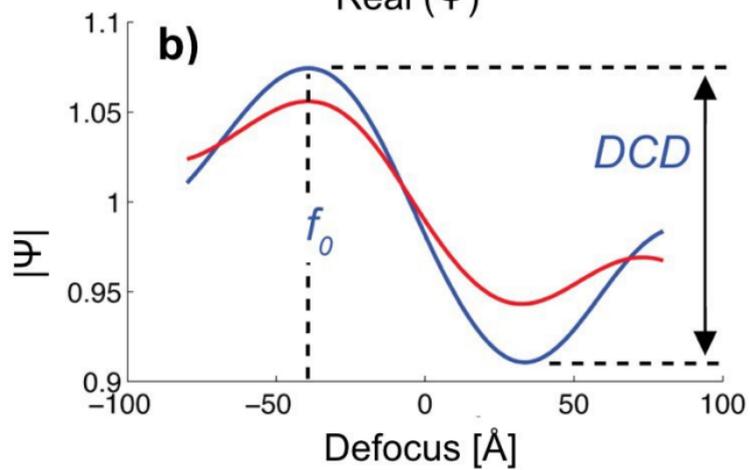
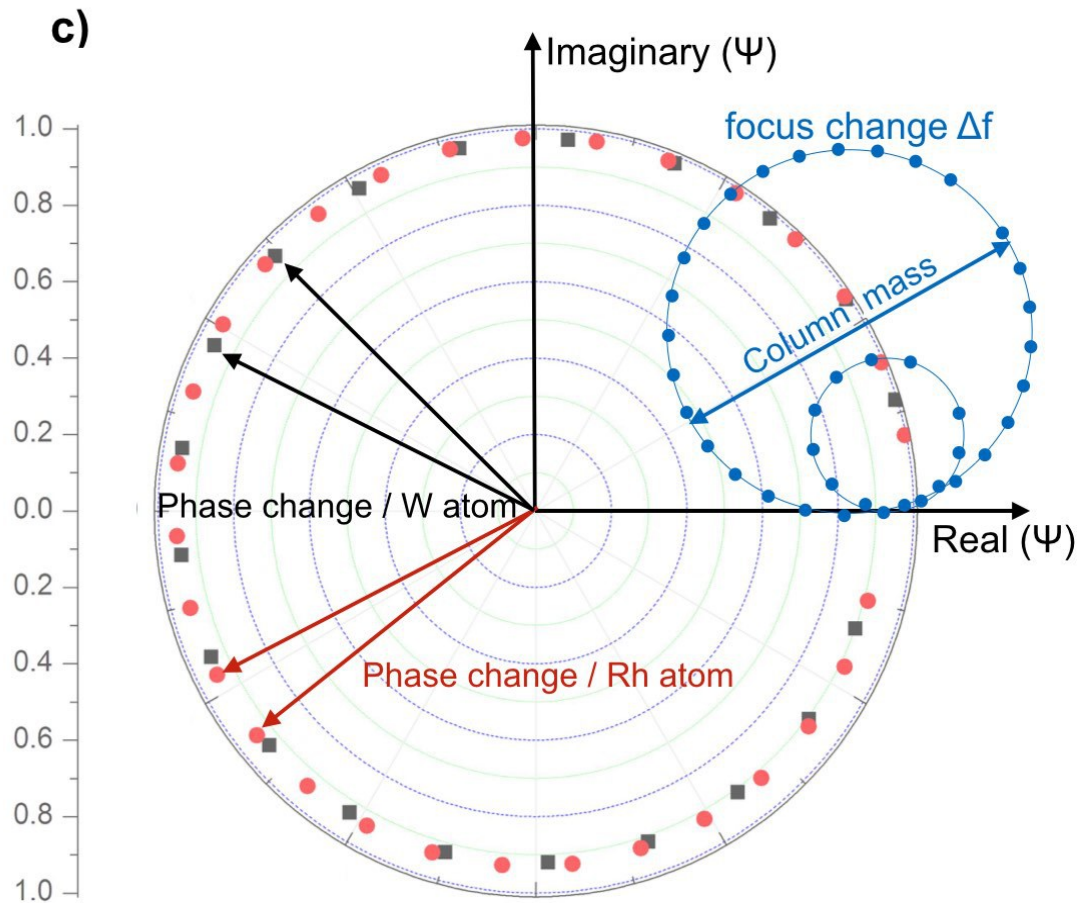

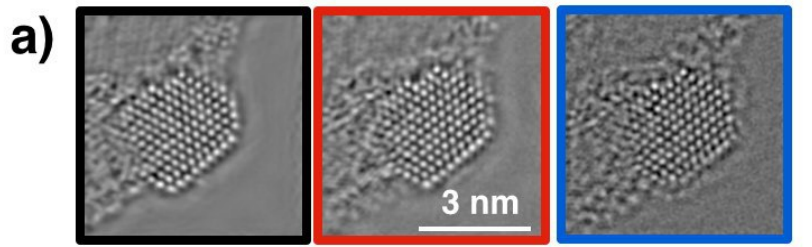
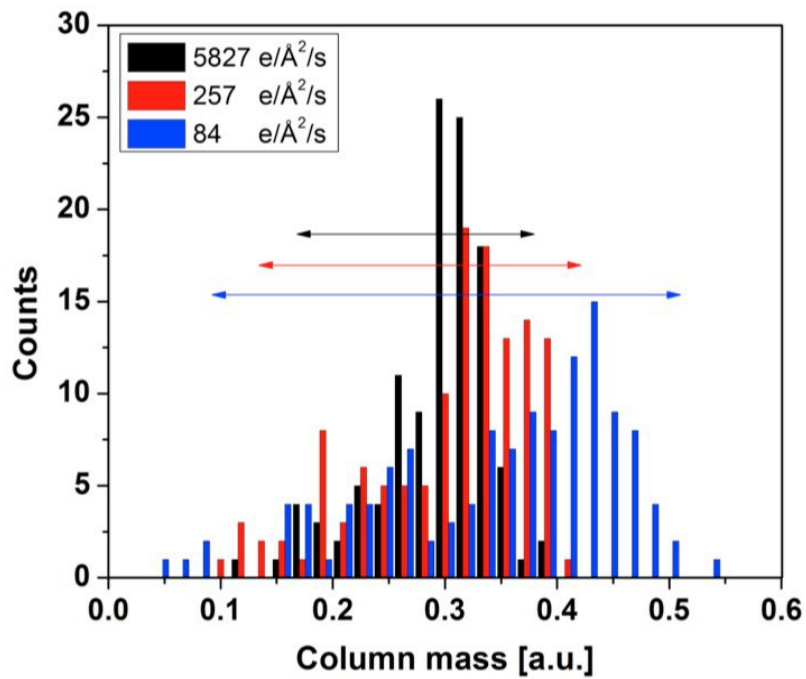
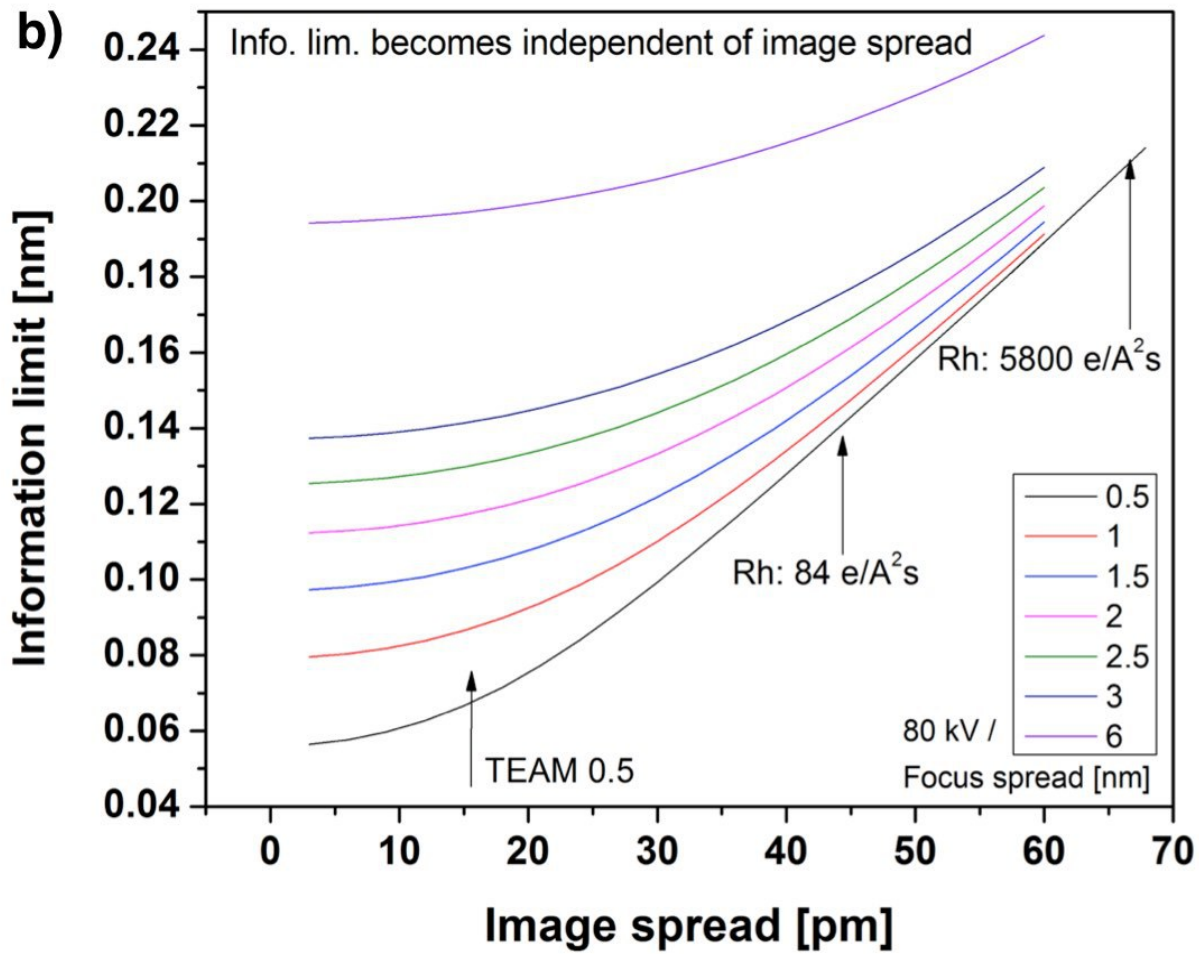

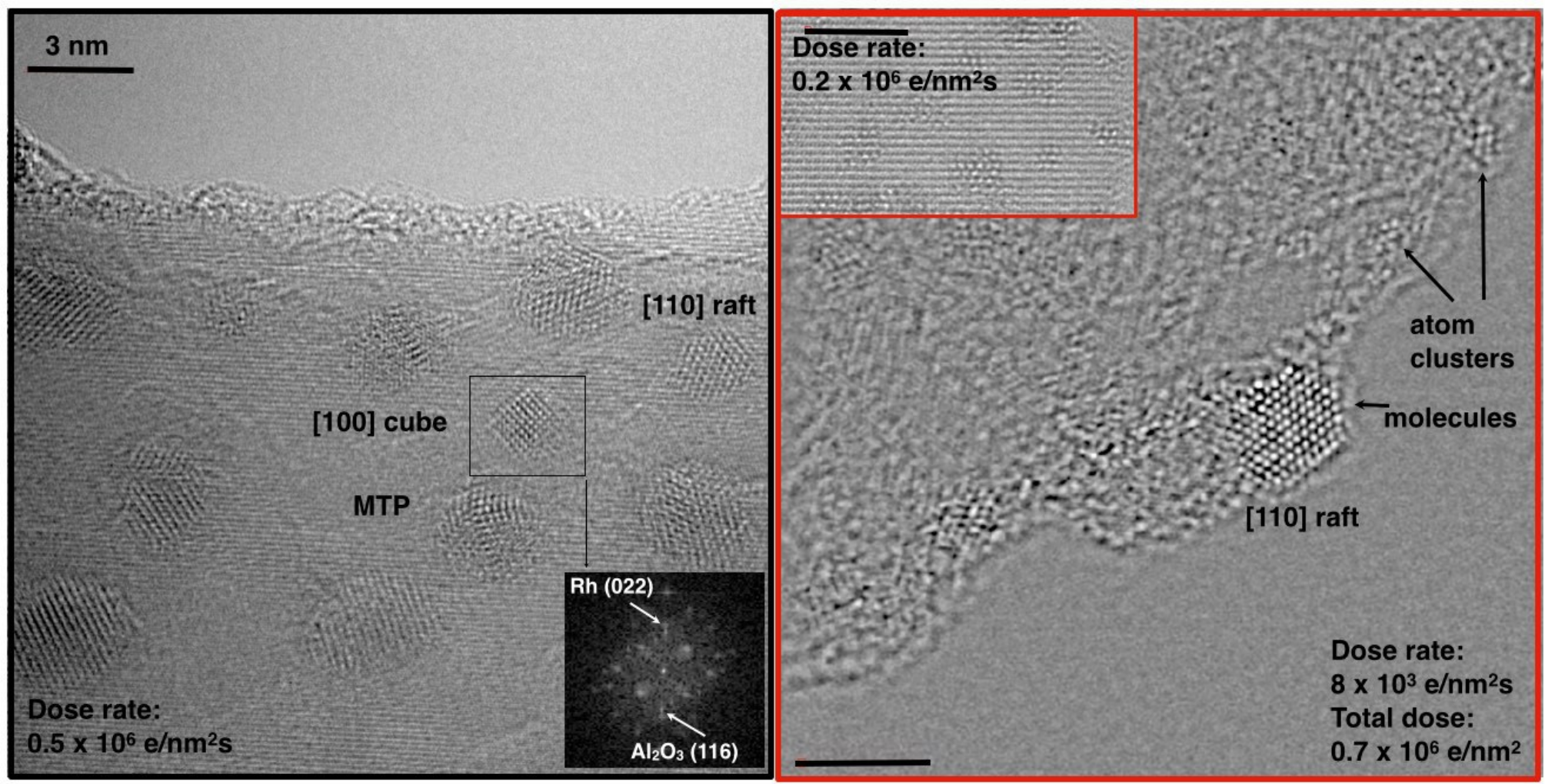

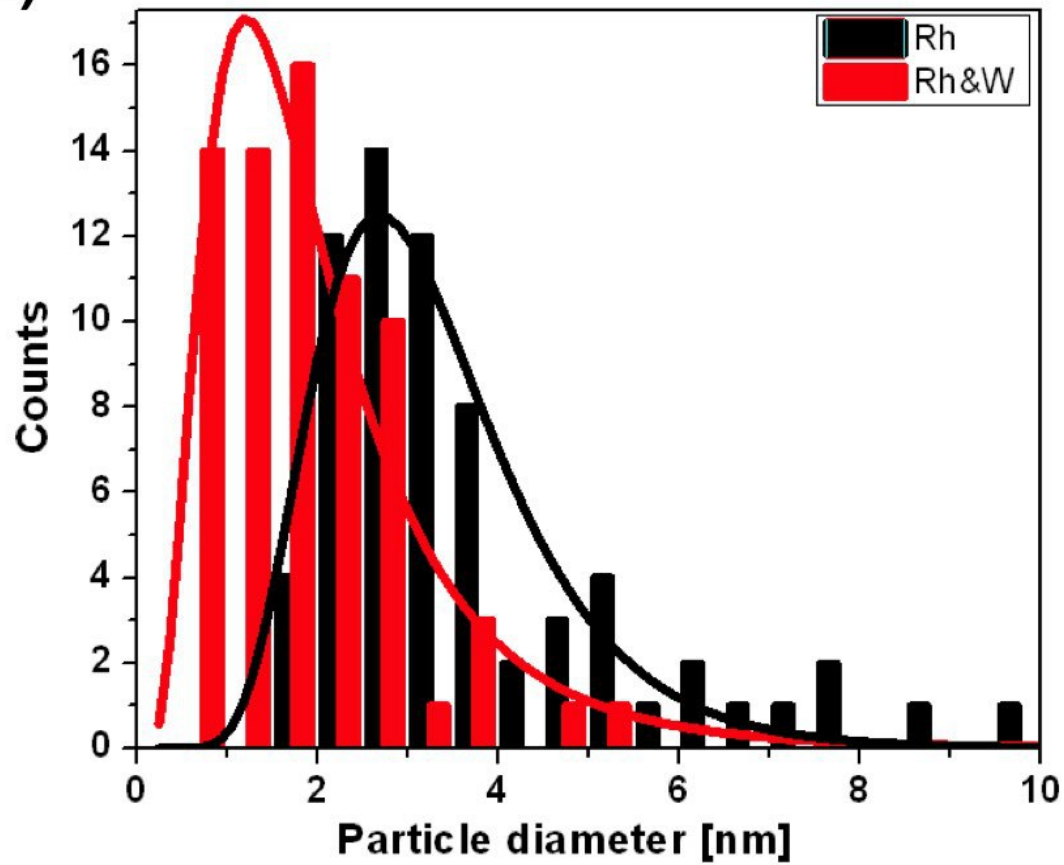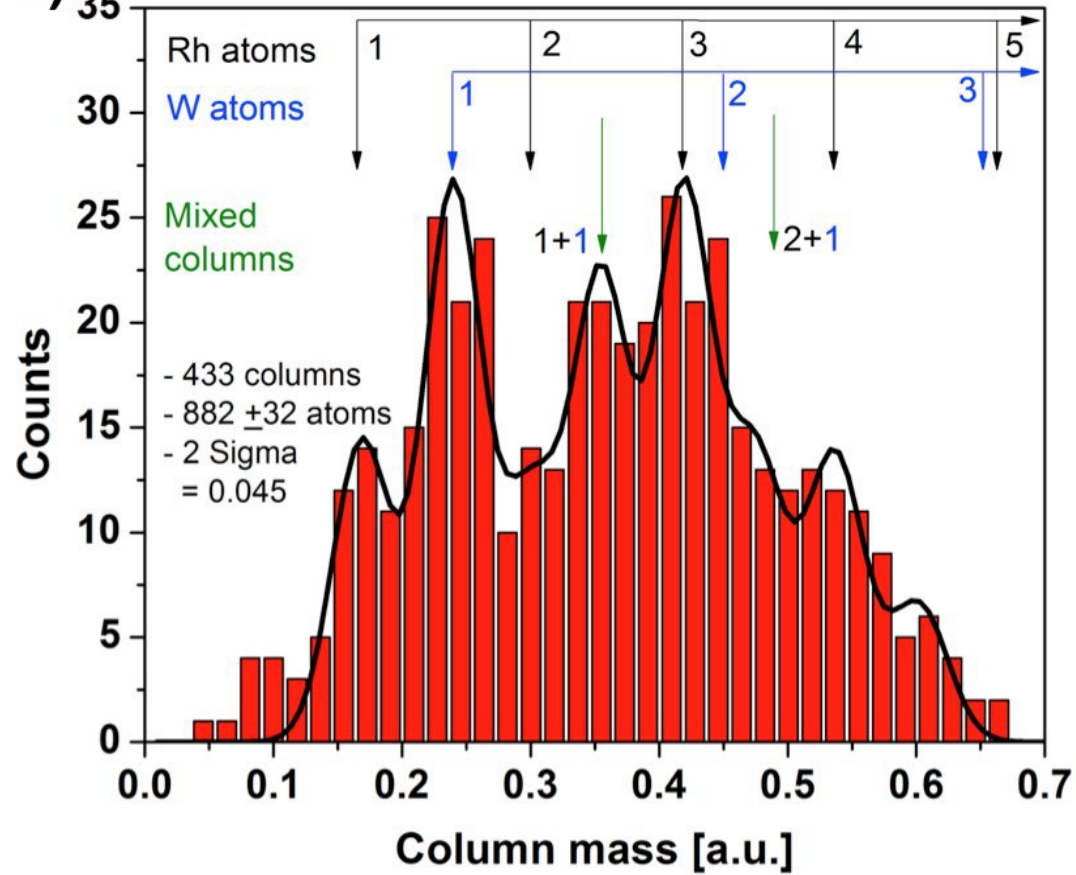

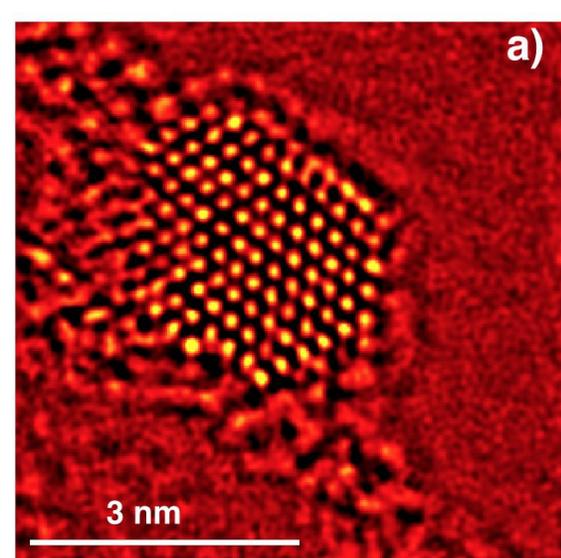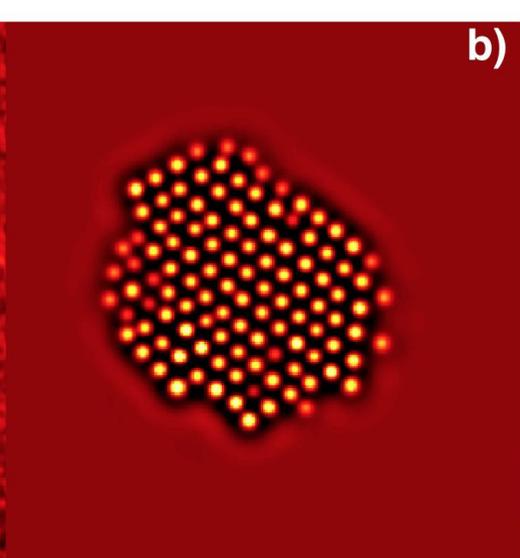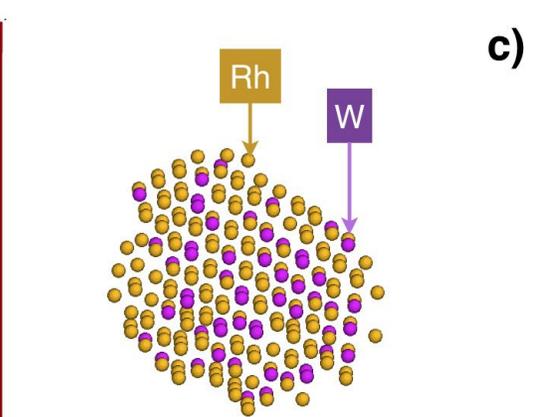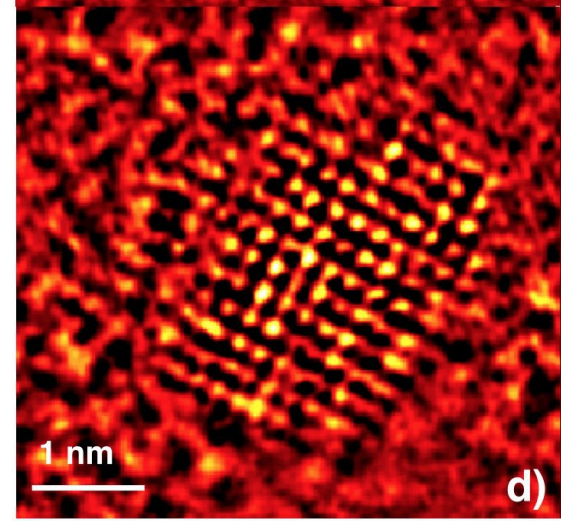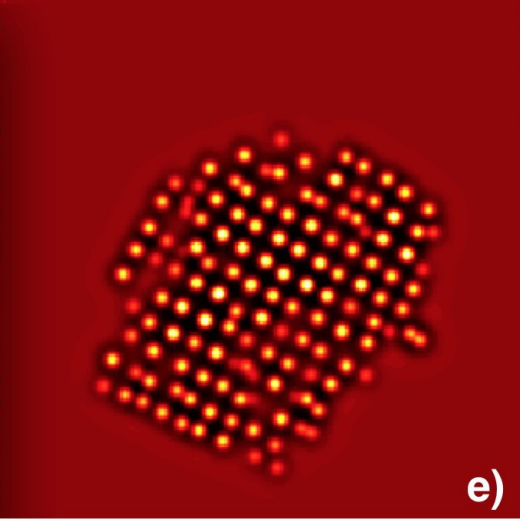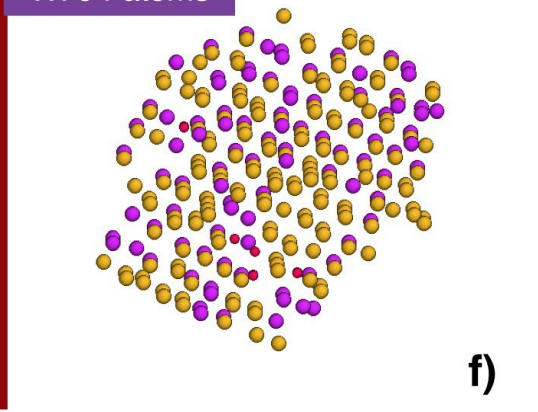

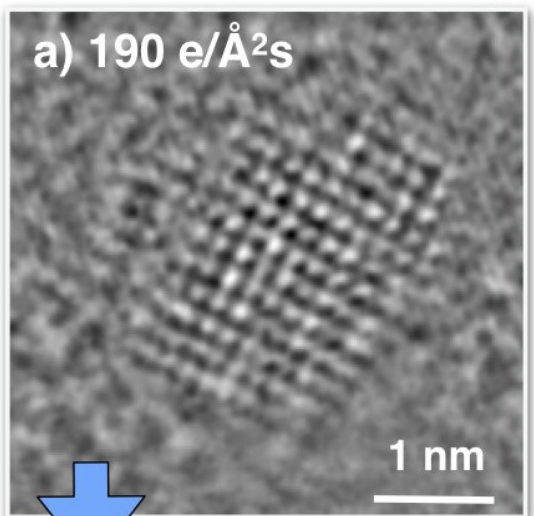
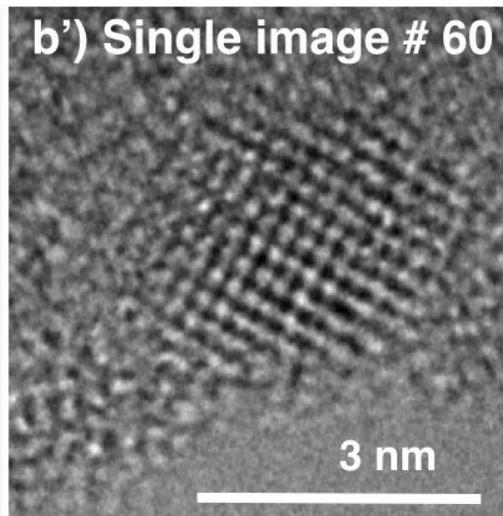
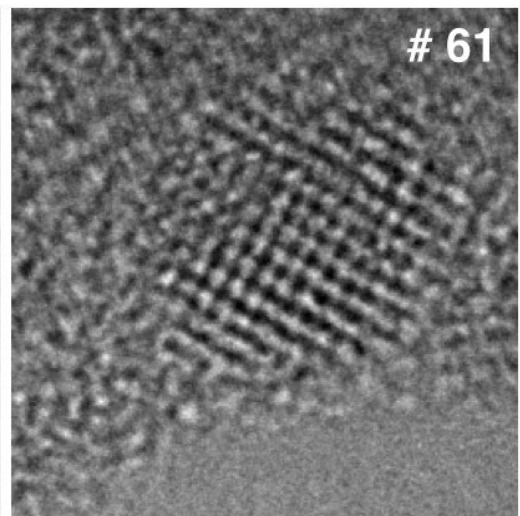
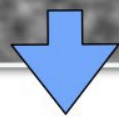
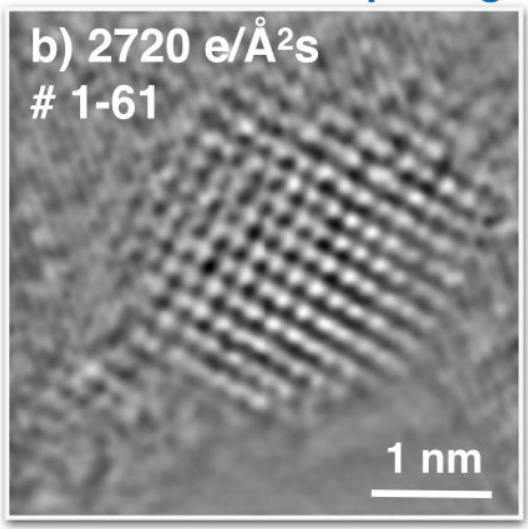
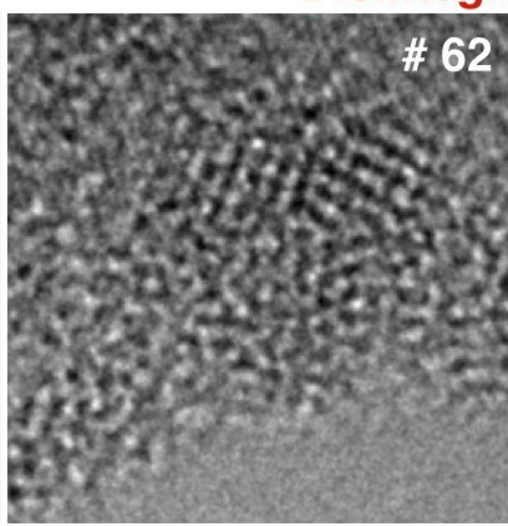
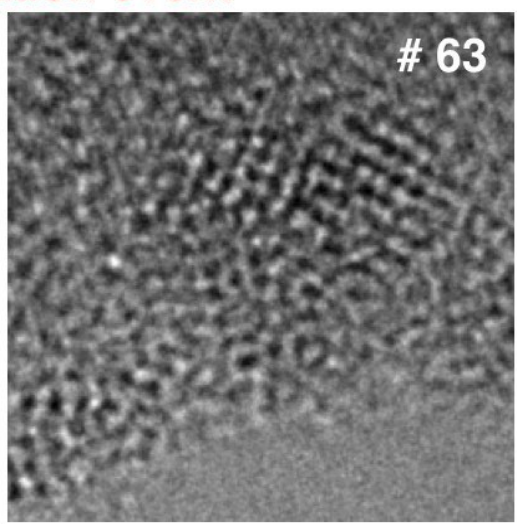
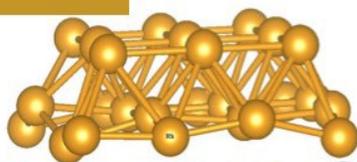
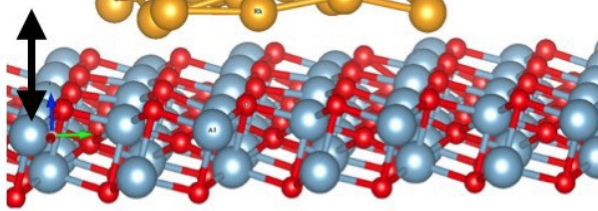
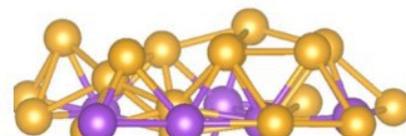
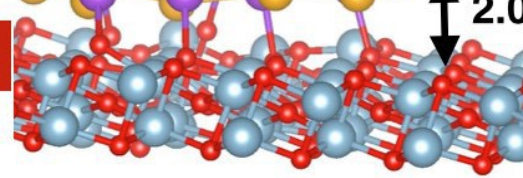

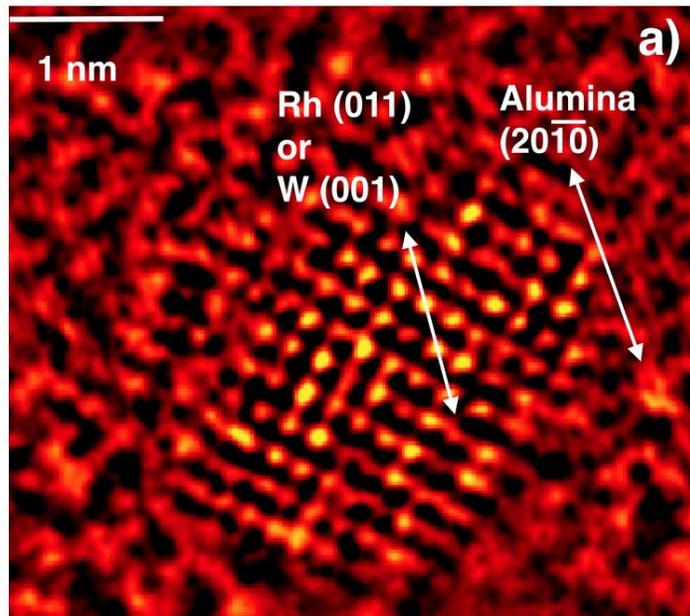
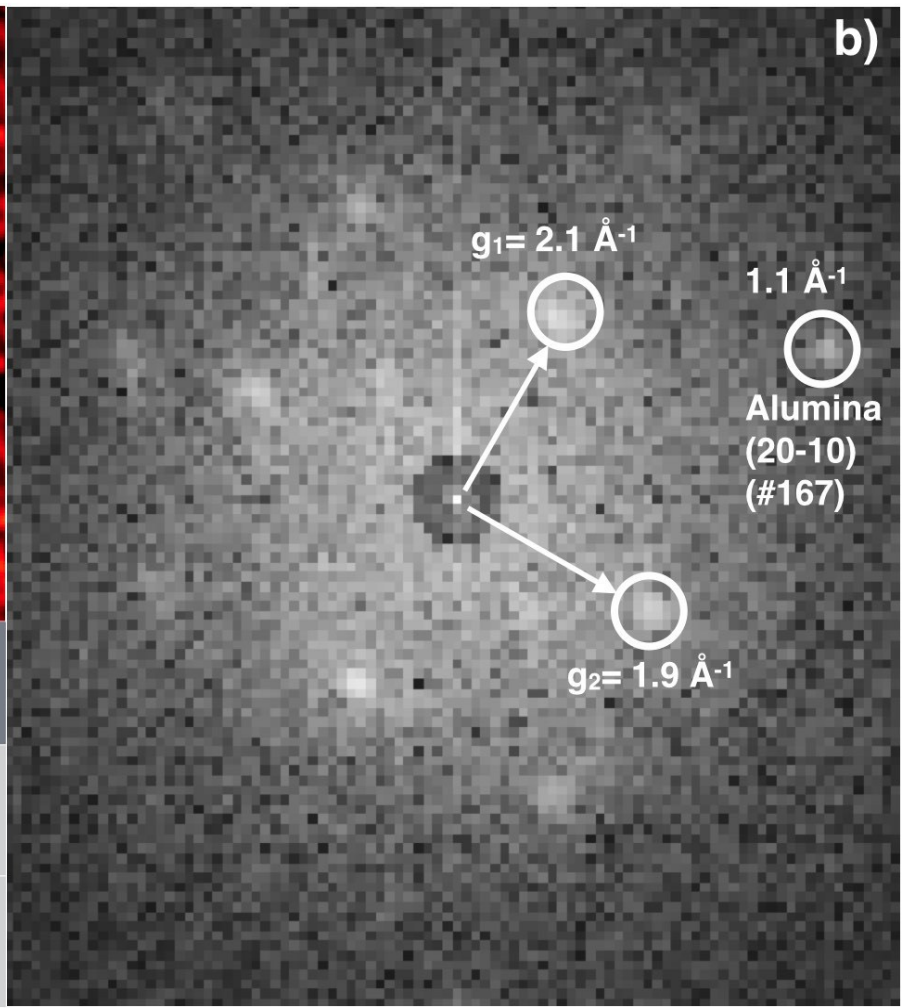

| [Å⁻¹] | Particle meas. | Tungsten [100] - cal. | Rhodium [100] - cal. |
|---|---|---|---|
| $g_1$ | 1/2.1 | 1/2.0 (0-11) | 1/1.9 (002) |
| $g_2$ | 1/1.9 | 1/2.0 (01-1) | 1/1.9 (020) |